\documentclass[showpacs,reprint,showkeys,preprintnumbers,amsmath,amssymb,aip,apl]{revtex4-1}
\usepackage[dvips]{graphicx}
\usepackage[dvips]{graphics}
\usepackage{dcolumn}
\usepackage{bm}
\usepackage[mathlines]{lineno}
\usepackage{siunitx}
\usepackage{tabularx}
\usepackage{multirow}
\usepackage{placeins}

\begin{document}
\title{Spin reorientations and crystal field modification in Ho$_{1-y}$Gd$_y$Al$_2$ compounds}\label{title}
\author{V. S. R. de Sousa}
\email{vinidesousa@gmail.com}
\affiliation{%
Instituto de F\'isica Armando Dias Tavares, Universidade do Estado do Rio de Janeiro (UERJ), 20550-013, Rio de Janeiro - RJ, Brazil. \\
}%
\author{L. E. L. Silva}
\author{A. M. Gomes}
\affiliation{%
Instituto de F\'isica , Universidade Federal do Rio de Janeiro (UFRJ), 20550-013, Rio de Janeiro - RJ, Brazil. \\
}%
\author{P. J. von Ranke}
\affiliation{%
Instituto de F\'isica Armando Dias Tavares, Universidade do Estado do Rio de Janeiro (UERJ), 20550-013, Rio de Janeiro - RJ, Brazil. \\
}%
%
%
\begin{abstract}
We have performed an experimental and theoretical investigation on the spin reorientation transitions Ho$_{1-y}$Gd$_y$Al$_2$ compounds. Crystallographic, magnetic and calorimetric measurements have been performed for five selected samples in the Ho$_{1-y}$Gd$_y$Al$_2$ series ($y$ = 0, 0.25, 0.5, 0.75 and 1.0). From the results, a linear increase in the lattice parameter as a function of Gd-content ($x$) is observed. Magnetization measurements show that all samples present collinear ferromagnetism. Specific heat measurements were performed in order to detail the first-order spin reorientation transition (FOSRT) that appears for HoAl$_2$ around 20 K. Furthermore, we investigated how the increase in Gd content affects the character of such phase transition. Our results show a slightly modification in the spin reorientation transition temperature, which moves to lower temperatures as Gd is increased in HoAl$_2$, whereas for the selected concentrations up to x = 0.75 the first-order character of the transition is maintained. Such behavior may lead to a large refrigerant capacity (RC) over a wide temperature range, which is an important parameter to evaluate the magnetocaloric effect (MCE) of such compounds. The theoretical analysis by means of a model Hamiltonian that includes crystal field, exchange and Zeeman interactions reveals a modification in the crystal-field acting on Ho ions as Gd-content is increased.
\end{abstract}
\maketitle
%
%
	The subject of caloric effects has attracted much attention from the scientific community in recent years. Within such effects, the magnetocaloric effect (MCE) and the barocaloric effect (BCE) are directly connected to the magnetic solid state refrigeration, which is supposed to be cleaner, more efficient and environmental friendly than the establish gas-compression based refrigeration. The MCE, which was discovered by Weiss and Piccard in 1917 (see Refs. [\onlinecite{weiss,smith}]), is a thermal response of magnetic materials when submitted to a magnetic field variation. The BCE is also related to a thermal response, but it occurs when such materials are subjected to pressure variations. On either case, this thermal response emerges as a result of intricate microscopic magnetic interactions, such as exchange, crystal field, Zeeman and magnetoelastic ones in consonance with phonon interactions, and is characterized by the so called \textit{adiabatic temperature change} ($\Delta T_{ad}$). Another important parameter to quantify the MCE and BCE is the \textit{isothermal entropy change} ($\Delta S_T$). Both $\Delta T_{ad}$ and $\Delta S_T$ are evaluated as a function of temperature under the variation of an external parameter: magnetic field for MCE and pressure for BCE (for a thorough discussion on the basics of the MCE and BCE the reader is referred to Refs. [\onlinecite{smith,tishin,gsch1,gsch2,bruck1,nilson2010}]).
	
	One of the key aspects of the MCE relies on the choice of the materials presenting large magnetocaloric effects. In rare earth based materials presenting anisotropy, it has been shown that one can define an anisotropic magnetocaloric effect (AMCE)\cite{tishin,vonranke2007}. Such effect may lead to large values of $\Delta T_{ad}$ and $\Delta S_T$, as in DyAl$_2$\cite{vonranke2009} and NdCo$_5$\citep{nikitin2010}, and differs from the usual MCE in one aspect related to the variation of the magnetic field. In the AMCE the intensity of the field is maintained fixed, meanwhile its orientation in relation to the crystalline axes is changed. The peak values for the AMCE is achieved around the so-called spin reorientation transition temperature (T$_{SR}$). A spin reorientation transition (SRT) may occur spontaneously, as a consequence of a change on the easy direction of magnetization as a function only of temperature. Or it may be induced by the application of a magnetic field or pressure (A good review on spin reorientations in rare earth magnets is given by Belov \textit{et al}. in Ref. [\onlinecite{belov}].). The case of first-order spin reorientation transitions (FOSRT's) is very interesting from the application point of view because it may lead to large values of the refrigerant capacity (RC) as, for instance, in Ho$_2$In\cite{zhang}, ErGa\cite{chen}, TbZn\cite{desousa2010} and HoZn\cite{desousa2011}.
	
	One may also be interested in cases that a FORST occurs when it is combined with a relatively high Curie temperature. For instance, in the RAl$_2$ series, HoAl$_2$ presents a FOSRT around 20 K, at this temperature the easy magnetization axis changes from $\langle{110}\rangle$ to $\langle{100}\rangle$ as temperature increases. Its Curie temperature (T$_C$=31 K), however, lies close to the spin reorientation temperature, therefore one can take advantage of such combination only at cryogenic temperatures. It would be interesting to tune the Curie temperature of HoAl$_2$ to higher temperatures, while maintaining the first order character of the spin reorientation temperature.
	
	In the 1960's, Swift and Wallace\cite{swift} showed that if one substitutes Gd for Ho in HoAl$_2$ host the critical temperature increases. In fact, the isomorphic pseudobinary alloys (Ho,Gd)Al$_2$ present a ferromagnetic coupling between Ho and Gd ions, such that one verifies a linear increase in the Curie temperature as Gd content is added. Nevertheless none information is given regarding the spin reorientation transition. In this work we exploit how doping Gd in HoAl$_2$ host influences the spin reorientation. 
		
	To this end, polycrystalline samples of Ho$_{1-y}$Gd$_y$Al$_2$, with $y$ = 0, 0.25, 0.5, 0.75 and 1.0, were prepared by arc-melting the elements in a high purity argon atmosphere on a water cooled copper hearth. The melting process was repeated four times in order to assure homogenization. The purity of the starting materials was 99.99 wt.\% for aluminum (Sigma Aldrich) and 99.9\% for the rare earth metals (Sigma Aldrich). The as-cast samples were annealed under argon atmosphere in a quartz tube at 1273 K for 19 hours to reduce stress and the presence of other phases. The samples were characterized by X-Ray diffraction (@ 300 K). Specific heat and magnetization measurements were performed in a commercial Proper Physical Measurement System (PPMS, Quantum Design). Magnetization data were collected either under ZFC and FC protocols @ H=200 Oe, in order to determine the critical temperatures. Specific heat data were collected @ zero magnetic field.

	In order to understanding the behavior of the samples, we make use of a model Hamiltonian which takes into account exchange, Zeeman and crystal field interactions. In this way, we can write the Hamiltonian acting on Gd and Ho ions as:
\begin{equation}
	{\cal H} = -\sum_{ij}{{\cal J}_{ij}\vec J_i \cdot \vec J_j}-\sum_i{g_i}\mu_B\mu_0 \vec H \cdot \vec J_i + {\cal H}_{CF}\label{eq:hmag},
\end{equation}
	
	The symbols in relation (\ref{eq:hmag}) have their usual meaning. ${\cal J}_{ij}$ is the exchange interaction between 4f moments (Gd and Ho), $\vec J$ the total angular momentum. ${\cal H}_{CF}$ represents the crystal field (CF) produced by neighboring ions at rare earth sites. The pseudo-binaries here studied crystallize in the cubic C15 Laves phase structure, in which the rare earth site is described by the point group T$_\text{d}$.\cite{johnston} Therefore the CF Hamiltonian presents only fourth and sixth order terms, and can be written in the Lea-Leask-Wolf notation as\cite{LLW}:
	\begin{equation}
		{\cal H}_{CF} = W\left[\frac{x}{F_4}\left(O_4^0 + 5O_4^4\right) \right. + 
		\left.\frac{1 - \left|x\right|}{F_6}\left(O_6^0 - 21O_6^4\right)\right]\label{eq:hcf}.
	\end{equation}
 
	Within the mean-field approximation, Hamiltonian (1) can be split into two contributions, one due to Gd-moments and the other due to Ho-moments. The effective fields acting on each sublattice are:
\begin{equation}
	{\vec H}^{eff}_{Ho} = \mu_0\vec H + (1-y){\cal J}_{HoHo}\langle\vec J_{Ho}\rangle + y{\cal J}_{HoGd}\langle\vec J_{Gd}\rangle,\label{eq:heff1}
\end{equation}
\begin{equation}
	{\vec H}^{eff}_{Gd} = \mu_0\vec H + (1-y){\cal J}_{HoGd}\langle\vec J_{Ho}\rangle + y{\cal J}_{GdGd}\langle\vec J_{Gd}\rangle.\label{eq:heff2}
\end{equation}

	Details on the calculations can be found in section 7.1 of Ref.[\onlinecite{nilson2010}].
	
	X-ray patterns (not shown here) reveal that all samples are single phase. From the Rietveld's refinement all compounds crystallized in the cubic C15 Laves phase structure. The lattice parameters are listed in Table~\ref{tab:lattice}.The variation of the lattice parameter across the series is linear. The results for HoAl$_2$ and GdAl$_2$ compounds are in good agreement with those reported in the literature\cite{buschow}. 

\begin{table}[h]
\caption{\label{tab:lattice} Lattice parameters of (Ho,Gd)Al$_2$ solid solutions.}
\begin{tabularx}{5cm}{p{2.5cm} c}\hline
y 		& a (\si{\angstrom})\\ \hline
0		& 7.8199	\\
0.25	& 7.8401	\\
0.5		& 7.8628	\\
0.75	& 7.8839	\\
1		& 7.9048	\\\hline
\end{tabularx}
\end{table}	


	In order to apply the model to the series of compounds Ho$_{1-y}$Gd$_y$Al$_2$ we considered the following set of molecular field parameters ${\cal J}_{HoHo}$ = 0.12 meV, ${\cal J}_{Gd}$ = 2.77 meV and ${\cal J}_{HoGd}$ = 0.56 meV. These parameters were chosen to adjust the Curie temperatures of the compounds, which were expected to increase linearly within the series. Crystal field parameters were chosen to account for the spin reorientation behavior (as will be discussed later in this paper). The critical temperatures (T$_C$/T$_{SR}$) obtained experimentally and crystal field parameters are summarized in Table~\ref{tab:tcs}. The Land\'e factors and total angular moment were obtained from Hund's rule.

\begin{table}[h]
\begin{ruledtabular}
\caption{\label{tab:tcs} Curie and spin reorientation temperatures, and crystal field parameters of the studied compounds.}
\begin{tabular}{l c c c c}
Compound  					 & T$_\text{C}$ (K) & T$_\text{SR}$ (K) & x & W (meV)\\\hline
HoAl$_2$					 & 		29.5		&       20.0	& -0.34  & 0.015 \\
Ho$_{0.75}$Gd$_{0.25}$Al$_2$ & 		67.3		&		18.4	& -0.36 & 0.012 \\
Ho$_{0.5}$Gd$_{0.5}$Al$_2$	 &	   105.9		&       17.4	& -0.37 & 0.0096  \\
Ho$_{0.25}$Gd$_{0.75}$Al$_2$ & 	   139.6		&       16.3	& -0.38 & 0.0082 \\
GdAl$_2$					 & 	   168.5		& 		-		& - &  -     \\
\end{tabular}
\end{ruledtabular}
\end{table}

\begin{figure*}
\centering
\includegraphics[width=\linewidth,angle=0]{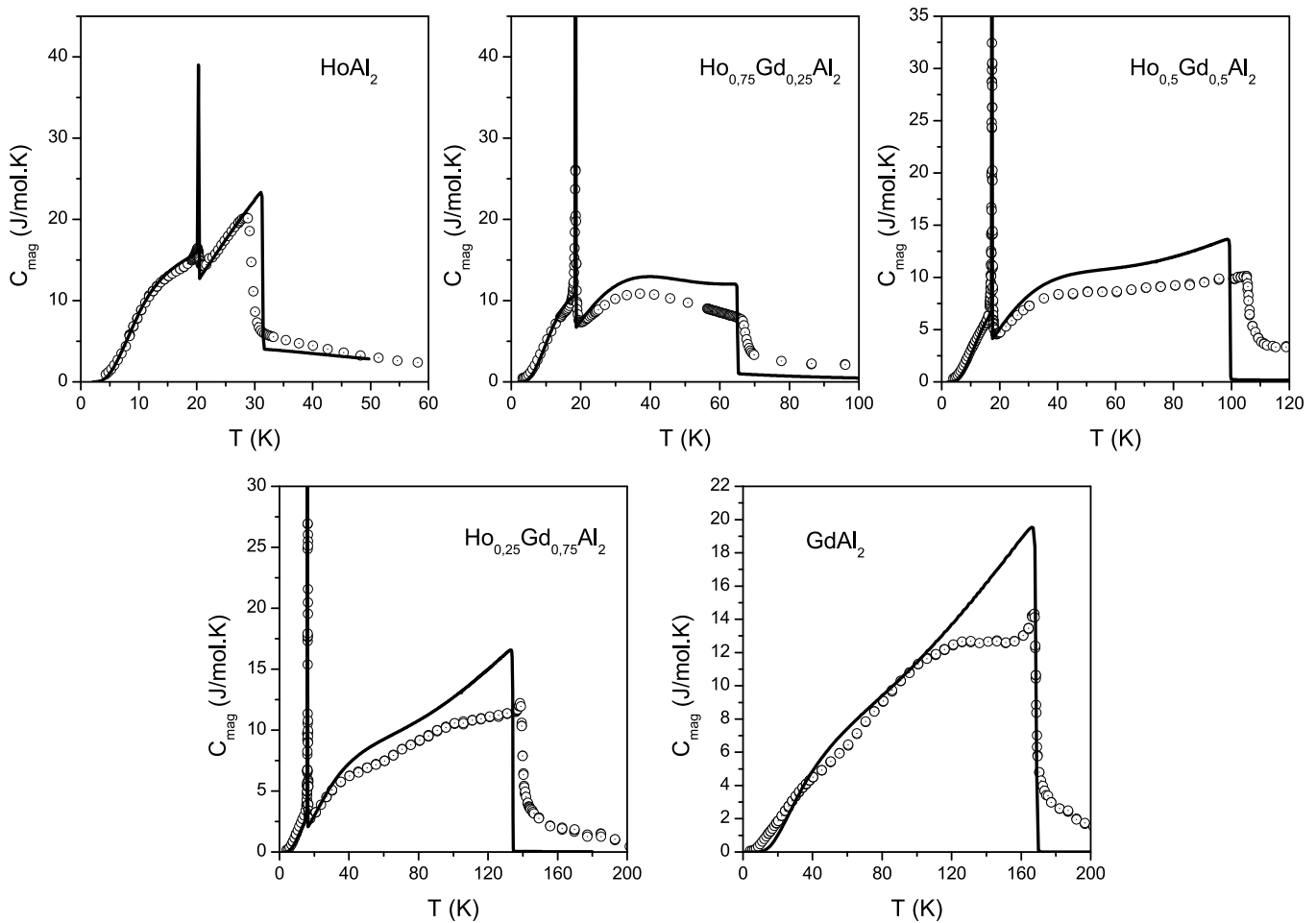}
\caption{\label{fig:cesp} Zero field magnetic specific heat for the Ho$_{1-y}$Gd$_{y}$Al$_2$ compounds. Open symbols represent experimental data and solid lines represent calculations.}
\end{figure*}

\begin{figure}[htb!]
\centering
\includegraphics[width=\linewidth]{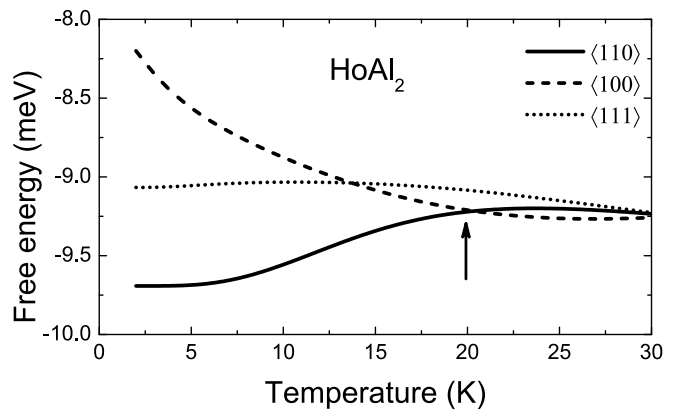}
\caption{\label{fig:fenergy} HoAl$_2$ zero field free energy calculated at the major cubic directions.}
\end{figure}

	Figure~\ref{fig:cesp} shows the calculated (solid lines) and measured (symbols) magnetic specific heat of Ho$_{1-y}$Gd$_y$Al$_2$ compounds at zero magnetic field. The specific heat of the compounds with Ho present two anomalies. The first one around 20 K corresponds to a first order spin reorientation transition (FOSRT) due to the competition between crystal field and molecular field interactions. In this FOSRT the magnetization spontaneously changes its direction from $\langle 110 \rangle$ towards $\langle 100 \rangle$ axis, as temperature is increased. In Fig.~\ref{fig:fenergy} we compare the free energy $F=-k_BT \ln Z$ of HoAl$_2$ calculated @ null field considering the magnetization along $\langle 110 \rangle$ (solid line), $\langle 100 \rangle$ (dashed line) and $\langle 111 \rangle$ (dotted line) directions. Note that as the temperature increases, a crossing occurs between the free energies at $\langle 110 \rangle$ and $\langle 100 \rangle$ directions (as indicated by the arrow). At temperatures above this crossing the $\langle 100 \rangle$ direction becomes the easy axis of magnetization.
	
	In the framework of the point charge model\cite{hutchings}, the crystal field acting on the rare ions in RAl$_2$ compounds is due mainly to aluminum ions, which are the nearest-neighbors to the rare earths in the crystal structure. We expected that the substitution of Ho for Gd would not lead to any considerable changes in the crystal field interaction. Therefore, the calculations were first carried considering the same set of crystal field parameters for the Ho$^{3+}$ sublattice. Nevertheless, the results showed a mismatch when compared to experimental results.
	
	Experimental data show that T$_{SR}$ slightly decreases as Gd content increase (see table~\ref{tab:tcs}). Such behavior implies that the $\langle{100}\rangle$ axis becomes more stable at lower temperatures for larger effective fields, such as the one produced by Gd ions. One should note that Gd-ions do dot contribute directly to the anisotropic energy, actually the dilution should make the interaction between Ho-ions, responsible for the magnetic anisotropy, less effective. As a consequence the thermal energy necessary to drive the magnetic moments from the $\langle 110 \rangle$ direction towards the $\langle 100 \rangle$ one is favored by Gd-content. This results in the lowering of T$_{SR}$ as Gd replaces Ho in HoAl$_2$.
	
	Keeping the same set of crystal field parameters in te calculations results in an opposite effect regarding T$_{SR}$, \textit{i.e.}, T$_{SR}$ would increase as Gd-content is increased. Therefore we decided to include a dependence of the crystal field parameters as a function of Gd-concentration, as listed in Table~\ref{tab:tcs}. Such a modification leads to the good agreement observed in Fig.~\ref{fig:cesp}. Furthermore, one should note that the systematic decrease of the CF-interaction, measured by the W parameter, is consistent with the expansion of the lattice caused by Gd-content.
	
\begin{figure}[htb!]
\centering
\includegraphics[width=\linewidth]{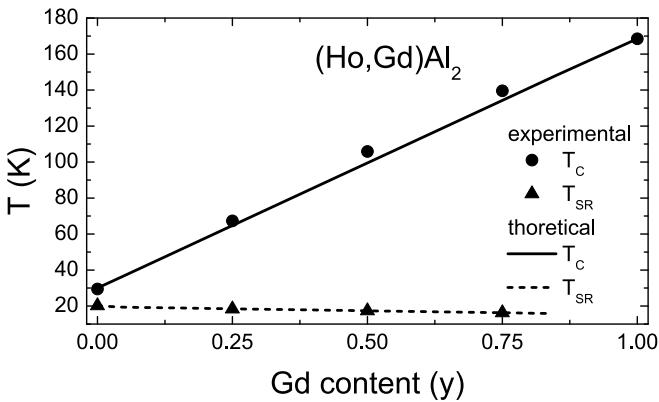}
\caption{\label{fig:diagram} Critical temperatures for (Ho,Gd)Al$_2$ compounds. The symbols represent experimental data and the lines theoretical results.}
\end{figure}

	Fig.~\ref{fig:diagram} shows how the critical and spontaneous reorientation temperatures varies along the (Ho,Gd)Al$_2$ series. The symbols represent experimental data obtained from magnetization/specific heat measurements and the lines (full and dotted) represent theoretical calculations. One should note the good agreement attained for the spin reorientation transition. The experimental critical temperature slightly departs from the linear behavior at intermediate Gd-concentrations. We have imposed in our calculations a linear variation of the critical temperature along the series, in order to obtain a variation proportional to the de Gennes factor:
\begin{equation}
	G = (1-y)(g_{Ho}-1)^2J_{Ho}(J_{Ho} + 1) + y(g_{Gd}-1)^2J_{Gd}(J_{Gd} + 1).
\end{equation}

	We have also verified the influence of an applied magnetic field on the spin reorientations. Depending on the intensity and direction of the magnetic field we observe different behaviors. The magnetic field may shift the reorientation transition to lower temperatures either keeping the first order character of the transition or changing it to second order. For more intense fields ($>$ 2T) the magnetization moves to the field direction even at the lowest temperature. Such behavior do not differ in quality from the ones reported in DyAl$_2$\cite{vonranke2007}, TbZn\cite{desousa2010} and HoZn\cite{desousa2011} by some of us.
\FloatBarrier

	The mechanism that lead to the modification of the crystal field acting in Ho-ions is not clear, although some possibilities include a change in the band structures of HoAl$_2$ and GdAl$_2$ (see Ref.[\onlinecite{teale1989}]). The model hamiltonian used to analyze the experimental results supports a modification in the crystal field caused mainly due to Gd ions.  We hope that the presented results may lead to further investigations that may clarify such behavior. It is worth noting that the presence of the two transitions (spin reorientation and ferro-paramagnetic transitions) may have a direct implication in the magnetocaloric and magnetotransport properties of the (Ho,Gd)Al$_2$ compounds.
\newline
%

	The authors would like to thank FAPERJ, Capes and CNPq for financial support.
\newline

\end{document}